\documentclass{icrc}

\usepackage{times}
 \usepackage[dvips]{graphicx} 

\begin{document}

\title{The CANGAROO-III Project: Status report}
\author[1]{M.~Mori}
\affil[1]{Insitute for Cosmic Ray Research, University of Tokyo,
Kashiwa, 277-8582 Chiba, Japan}
\author[2]{A. Asahara} 
\affil[2]{Department of Physics, Kyoto University, Sakyo-ku, Kyoto 606-8502, Japan}
\author[3]{G.V. Bicknell}
\affil[3]{MSSSO, Australian National University, ACT 2611, Australia}
\author[4]{R.W. Clay}
\affil[4]{Department of Physics and Math.\ Physics, University of Adelaide, SA 5005, Australia}
\author[5]{P.G. Edwards}
\affil[5]{Institute of Space and Astronautical Science, Sagamihara, Kanagawa 229-8510, Japan}
\author[1]{R. Enomoto}
\author[6]{S. Gunji} 
\affil[6]{Department of Physics, Yamagata University, Yamagata, Yamagata 990-8560, Japan}
\author[7]{S. Hara} 
\affil[7]{Department of Physics, Tokyo Insitute of Technology, Meguro-ku, Tokyo 152-8551, Japan}
\author[8]{T. Hara}
\affil[8]{Faculty of Management Information, Yamanashi Gakuin University, Kofu, Yamanashi 400-8575, Japan}
\author[9]{S. Hayashi}
\affil[9]{Department of Physics, Konan University, Kobe, Hyogo 658-8501, Japan}
\author[10]{C. Itoh}
\affil[10]{Faculty of Science, Ibaraki University, Mito, Ibaraki 310-8512, Japan}
\author[1]{S. Kabuki}
\author[9]{F. Kajino}
\author[1]{H. Katagiri}
\author[1]{A. Kawachi} 
\author[11]{T. Kifune}
\affil[11]{Faculty of Engineering, Shinshu University, Nagano, Nagano 380-8553, Japan}
\author[2]{H. Kubo}
\author[7]{J. Kushida}
\author[9]{S. Maeda}
\author[9]{A. Maeshiro}
\author[12]{Y. Matsubara} 
\affil[12]{STE Laboratory, Nagoya University, Nagoya, Aichi 464-8601, Japan}
\author[13]{Y. Mizumoto} 
\affil[13]{National Astronomical Observatory of Japan, Mitaka, Tokyo 181-8588, Japan}
\author[14]{H. Muraishi}
\affil[14]{Ibaraki Prefectural University of Health Sciences,
 Ami, Ibaraki 300-0394, Japan}
\author[12]{Y. Muraki}
\author[8]{T. Naito} 
\author[15]{T. Nakase}
\affil[15]{Department of Physics, Tokai University, Hiratsuka, Kanagawa 259-1292, Japan}
\author[15]{K. Nishijima}
\author[1]{M. Ohishi}
\author[1]{K. Okumura}
\author[4]{J.R. Patterson}
\author[4]{R.J. Protheroe}
\author[7]{K. Sakurazawa}
\author[1]{R. Suzuki}
\author[4]{D.L. Swaby}
\author[2]{T. Tanimori}
\author[6]{F. Tokanai} 
\author[1]{K. Tsuchiya}
\author[1]{H. Tsunoo}
\author[15]{K. Uruma} 
\author[6]{A. Watanabe}
\author[10]{S. Yanagita}
\author[10]{T. Yoshida}
\author[16]{T. Yoshikoshi}
\affil[16]{Department of Physics, Osaka City University, Osaka, Osaka 558-8585, Japan}

\correspondence{M.~Mori (Email: {\tt morim@icrr.u-tokyo.ac.jp} and 
{\tt http://icrhp9.icrr.u-tokyo.ac.jp})}

\runninghead{Mori et al.\ : The CANGAROO-III Project}
\firstpage{1}
\pubyear{2001}

 \titleheight{11.5cm} 

\maketitle

\begin{abstract}
We report on the status of the construction of an array of four 10 m
atmospheric Cherenkov telescopes for gamma-ray astronomy, 
near Woomera, in South Australia -- the CANGAROO-III project. 
The first telescope of this array is the upgraded
version of the CANGAROO-II 7 m telescope and has been in operation since
March 2000. The second telescope, an improved version of the first, is
being constructed for installation in late 2001. 
Stereoscopic observation of sub
TeV gamma-rays with the two 10 m telescopes will begin in 2002 and the full
array will be operational in 2004.
\end{abstract}

\section{Introduction}

Following the successful operation of the 
CANGAROO-I 3.8m telescope and the CANGAROO-II 7m 
telescope \citep{Tan99} in Woomera, South Australia
(136$^\circ$47$'$E, 31$^\circ$06$'$E, 160m a.s.l.), 
we have started the construction of an array
of four 10m telescopes, CANGAROO-III, to explore the sub-TeV
gamma-ray sky in the southern hemisphere with high accuracy
by utilizing stereoscopic observations of Cherenkov light images
\citep{Mor00, Mor01, Tan01, Eno01a}.
The 
full array will be operational in 2004.

In this paper we report on the expansion of the 7m telescope
into the first 10m telescope in 2000 and the work in progress for 
construction of three more telescopes.

\section{The First 10m Telescope}

The first telescope of this array,
which has been in operation since April 2000, 
is the upgraded version of the CANGAROO-II 7 m telescope 
The 7m telescope is described in \citet{Tan99} and so here
we describe only more recent improvements.
Table \ref{7to10m} summarizes the properties of the 7m and
the 10m telescopes.

\begin{table}
\begin{tabular}{lcc} \hline
 & 7m telescope & 10m telescope \\ \hline
Focal length & 8m & 8m \\
80cm CFRP mirrors & 60 (30m$^2$) & 114 (57m$^2$) \\
Number of PMTs & 512 (1/2") & 552 (1/2") \\
Readout & TDC & TDC \& ADC \\
Point image size & 0.15$^\circ$ (FWHM) & 0.20$^\circ$ (FWHM) \\
Operation & May '99--Feb '00 & Mar '00-- \\ \hline
\end{tabular}
\caption{Properties of the 7m and the 10m telescopes.}
\label{7to10m}
\end{table}

Results from observations with this telescope are reported
in these proceedings \citep{Eno01a,Har01,Kus01,Nis01,Oku01}.

\subsection{10m reflector}

\begin{figure}[t]
 \includegraphics[width=8.3cm]{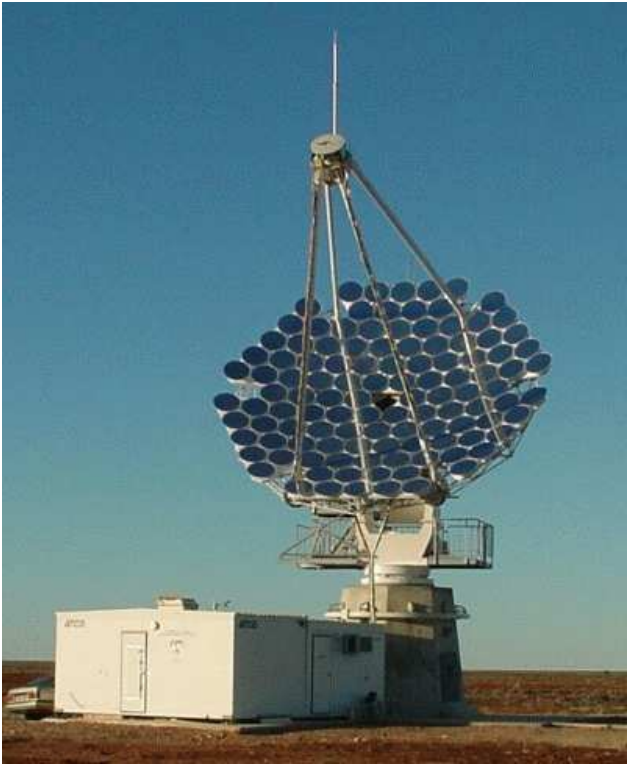} 
 \caption{The first 10m telescope in Woomera, South Australia. The huts
   contain the electronics and the telescope power.}
 \label{10m}
 \end{figure}

The 7m telescope was expanded to its full 10m diameter
by adding small spherical 80cm diameter mirrors 
made of CFRP \citep{Kaw01}.
This expansion was done in February 2000 as the first
step of the CANGAROO-III project (Fig.\ref{10m}). 
Now the reflector consists of 114 mirrors arranged
on a parabolic surface.
The point image size was measured by observing stars with
a CCD camera viewing the prime focal plane. 
It became a little worse, $0.20^\circ$ FWHM, 
than that of the 7m telescope, $0.15^\circ$ FWHM.
However, Monte Carlo simulation shows this has
small effect in analyzing Cherenkov images.

%

\subsection{552 channel camera}

\begin{figure}[t]
 \begin{center}
 \includegraphics[width=7cm]{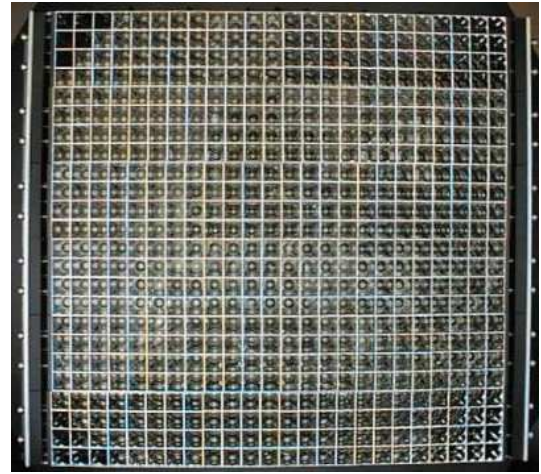} 
 \caption{The 552-channel imaging camera at the prime focus of the first
   10m telescope equipped with light guide in front of PMT's.}
 \label{552ch}
 \end{center}
 \end{figure}

The imaging camera has been upgraded to 552 pixels, each a
half-inch photomultiplier (PMT's, Hamamatsu R4124UV),
by adding 10 pixels at each of the four corners (Fig.\ref{552ch}).
The camera covers a field-of-view of about 3 degrees.

\subsection{Electronics}

The electronics for the 7m telescope is described in \citet{Mor99}.
At the same time as the reflector upgrade, we made some improvement
for enhancing the dynamic range in the signal processing \citep{Kub01}.
Signals from the camera
are divided by new active buffers and fed to discriminator
modules and new ADC's.
The charge-sensitive ADC modules are housed in the VME-9U standard 
and have 32 channels of 12-bit resolution.
The internal delay of 150 ns for each channel makes external delay
cables unnecessary. The minimum gate width is 50 ns.
The online CPU has been changed from a Sun SPARC
(Solaris) to a Pentium (Linux) with a PCI-VME interface.
This reduced the system deadtime significantly.

\section{New 10m telescopes}

\begin{figure}[t]
 \includegraphics[width=8.3cm]{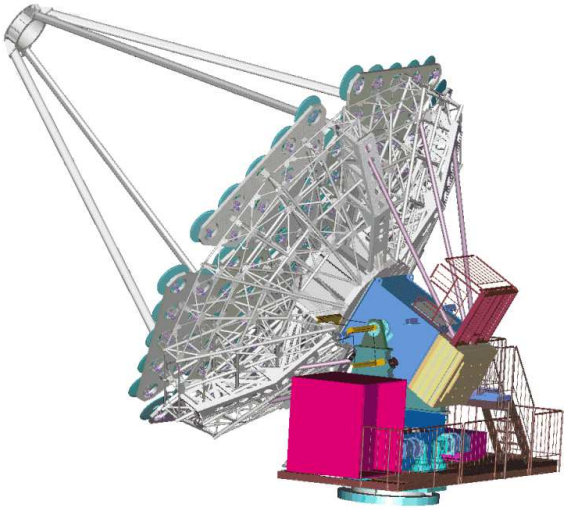} 
 \caption{A drawing of the second 10m telescope.
  The large box on the verandah is an electronics container.}
 \label{gammatel}
 \end{figure}

Four telescopes will be set at each corner of a diamond with sides of
about 100m. Performance of the system of telescopes is given
in detail elsewhere \citep{Eno01b}.
Here we describe the major improvements for the telescopes to be
installed in 2001 and later (Fig.\ref{gammatel}).

\subsection{Reflector}

\begin{figure}
 \includegraphics[width=8.3cm]{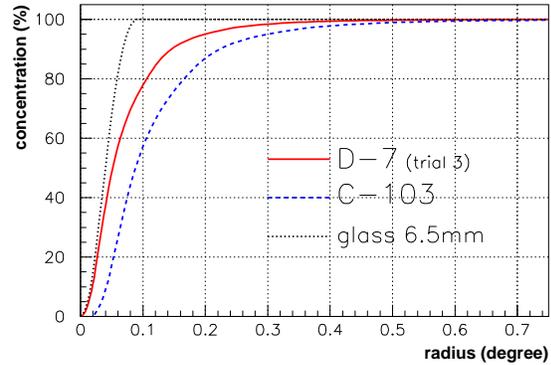} 
 \caption{Plot of light concentration within specific radius. ``D-7" is
  a new sample mirror and ``C-103" is a one used in the first 10m telescope.
  For comparison a plot of a sample glass mirror of 6.5mm thickness is 
  also plotted.}
 \label{conc}
 \end{figure}

The reflector design is the same as the first 10m telescope.
One hundred and fourteen CFRP mirrors will be arranged to a
parabola shape for each telescope (57 m$^2$).
The optical quality of mirrors has been improved (Fig.\ref{conc}) by
refining the production process.
The mirror attitude adjustment system has been redesigned to
match our needs and to save cost and weight.

\subsection{Telescope control}

Each telescope is controlled by a PC running a Linux operating system
with a realtime extension (KURT) in alt-azimuth manner.
A master PC issue directives to each control PC via
a network and tracking modes can be flexibly changed.
Clocks are synchronized by NTP software to a GPS receiver.

\subsection{Camera}

\begin{figure}
 \includegraphics[width=8.3cm]{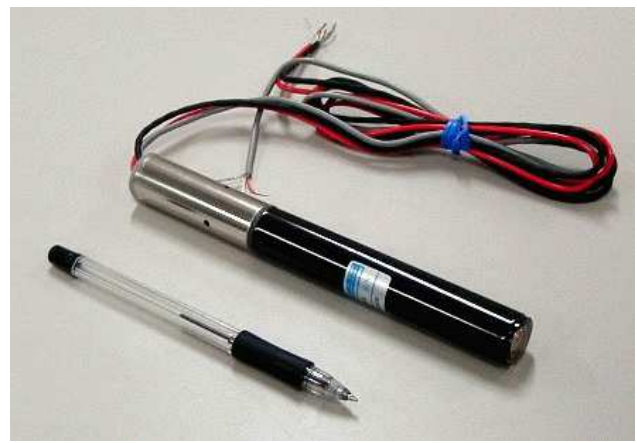} 
 \caption{New PMT assembly (3/4", Hamamatsu R3479UV) with a preamplifier inside.}
 \label{pmt}
 \end{figure}

\begin{figure}
 \includegraphics[width=8.3cm]{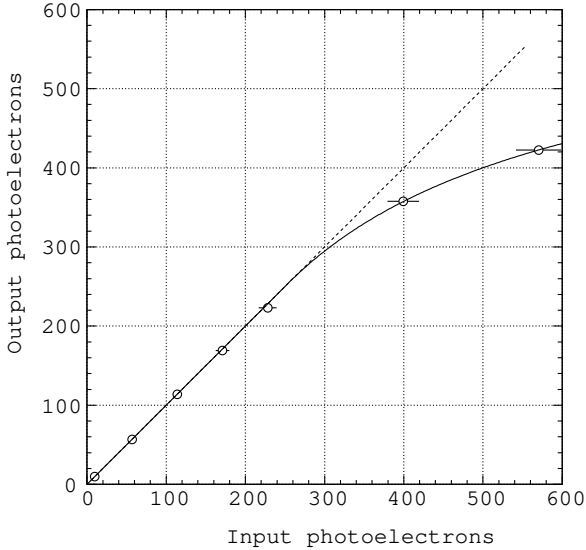} 
 \caption{Linearity of the new PMT assembly. The PMT gain was
  set to be $2\times10^5$ and the amplifier gain 60.
  One can see a good linearity up to about 300 photoelectrons.}
 \label{linearity}
 \end{figure}

\begin{figure}
 \includegraphics[width=8.3cm]{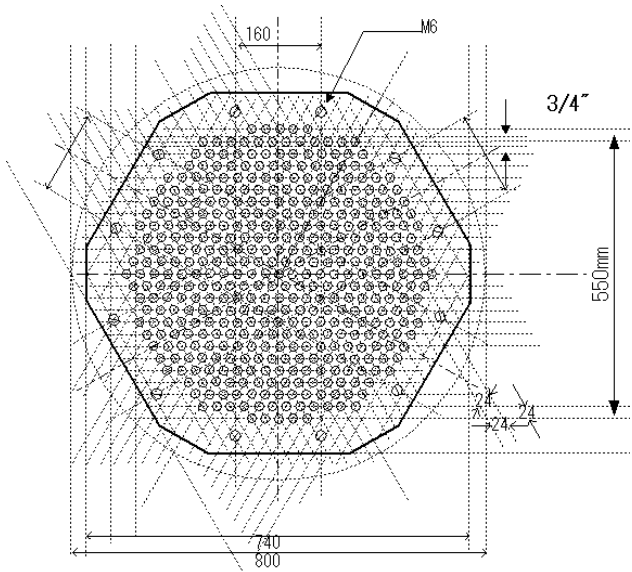} 
 \caption{Design of the new imaging camera. Each circle corresponds
  to an assembly of a 3/4" PMT in spacing of 24mm. The 427 pixel
  camera subtends a  field-of-view of about 4 degrees}
 \label{camera}
 \end{figure}

Imaging cameras have been redesigned to cover a wider field-of-view
at moderate cost. Hamamatsu R3479UV (3/4") PMT's are selected
and the base has been designed with a preamplifier inside
(Fig.\ref{pmt}).
A prototype of this assembly showed good pulse height resolution
and good linearity beyond the 300 photoelectron level (Fig.\ref{linearity}).
Four hundred and twenty seven PMT's are arranged in hexagonal 
shape with 24mm spacing
and cover a field-of-view of about 4 degrees (Fig.\ref{camera}).
High voltages for PMT's are generated in 
individually-controlled sources
(CAEN SY527 /A932) via a VME module (CAEN V288).
Positive voltages are supplied to avoid discharging problems
between light guides and photocathodes.
Each PMT has a newly designed light guide
in front of its photocathode thus reducing dead space between
photocathodes and collecting twice as many photons \citep{Kaj01}.

\subsection{Electronics}

The frontend module has been redesigned.
It is a VME-9U card and amplifies signals from PMT assemblies
and feeds to an ADC, discriminates them and feeds to
a TDC, an internal scaler and a trigger circuit.
The VME-based ADC has been improved for faster data transfer
compared with those used in the first 10m telescope.
VME TDCs (CAEN V673, multievent/multihit, 1~ns resolution) 
with CERN V430-type crates (with $-5.2$~V power).
PLD-based pattern trigger modules are under development in order
to reduce accidental triggers caused by nightsky background photons.
Faster data readout are achieved by use of a VME-based 
Pentium CPU running a Linux operating system.
All the electronics are set on the verandah of the telescope
base to save cable length and retain signal bandwidth 
(Fig.\ref{gammatel}.
Data from each telescopes are collected via network
and stored at a central disk.
Details are given in \citet{Kub01}.

\subsection{Monitors}

Cloud monitors detect infrared radiation from clouds making use
of a thermopile module and
supply useful information on data quality \citep{Cla98}.
Weather monitors can record temperature, humidity and wind
speed. These data are read out via serial line connection
and stored for offline analysis.
A blue LED light source driven by a fast pulser at the reflector pole 
works as a field-flattener.
Another LED source at a distant hill is used as a total
gain calibrator \citep{Pat01}.
CCD cameras will be equipped to monitor star fields and focal plane
images.

%


\section{Schedule}

The second 10m telescope will be installed in late 2001.
The first stereoscopic observations will start in early 2002.
The third telescope will be in place in 2002 and the fourth
in 2003. Then the full array will be operational
by early 2004.

\section{Summary}
The CANGAROO-III project to search for high-energy gamma-ray
objects in the southern hemisphere with an array of four
10m atmospheric Cherenkov imaging telescopes will be ready
in 2004. The final goal will be the energy threshold of
100 GeV
and the angular resolution of less than 0.1 degree.

\begin{acknowledgements}
We thank Communication Systems Center, Mitsubishi Electric 
Corporation, and DSC Woomera 
for their assistance in constructing the
telescopes. This project is supported by a Grant-in-Aid for
Scientific Research of Ministry of Education,
Culture, Science, Sports and technology of Japan,
and the Australian Research Council.
\end{acknowledgements}

\end{document}